\documentclass[aps,onecolumn,showpacs,footinbib]{revtex4}%
\usepackage[latin9]{inputenc}
\usepackage{amssymb}
\usepackage{amsmath}
\usepackage{amscd}
\usepackage{graphicx}
\usepackage{subfigure}
\usepackage{float}
\begin{document}
\draft
\title{Macroscopic displaced thermal field as the entanglement catalyst}
\author{Shi-Biao Zheng\thanks{%
E-mail: sbzheng@pub5.fz.fj.cn}}
\address{Department of Electronic Science and Applied Physics\\
Fuzhou University\\
Fuzhou 350002, P. R. China}
\date{\today }

\begin{abstract}
We show that entanglement of multiple atoms can arise via resonant
interaction with a displaced thermal field with a macroscopic photon-number.
The cavity field acts as the catalyst, which is disentangled with the atomic
system after the operation. Remarkably, the entanglement speed does not
decrease as the average photon-number of the mixed thermal state increases.
The atoms may evolve to a highly entangled state even when the photon-number
of the cavity mode approaches infinity.
\end{abstract}


\vskip 0.5cm \maketitle \narrowtext

In quantum mechanics, superposition effects give arise to many striking
features. Superpositions of product states of composite systems leads to
quantum entanglement, which is an entirely quantum-mechanical effect and
results in phenomena that can not be explained in classical terms. On one
hand, an entangled state of two or more particle reveals nonlocal structure
of quantum theory, providing a basis for the test of quantum mechanics
against local hidden variable theories [1,2]. On the other hand,
entanglement is an essential ingredient for quantum informatiom processing,
such as quantum cryptography [3] and teleportation [4]. The Jaynes-Cummings
model (JCM ) [5, 6], which describes the interaction of a two-level atom and
a single-mode electromagnetic field, is a typical system for producing
entanglement. It has been shown that for certain pure initial states,
entanglement between the atom and cavity mode oscillates with time [7]. The
cavity mode can also act as the catalyst for the synthesis of multiatom
entanglement [8]. Over the past few decades, there have been various
generalizations of the JCM. One of the typical examples is the
Tavis-Cummings model [9], describing the interaction of multiple two-level
atoms and the cavity mode.

Recent advances in microwave cavity QED techniques, with Rydberg atoms
interacting with a superconducting cavity, allow the test of many
interesting quantum effects arising from the interaction of atoms with a
quantized cavity field [8]. Up to now, entangled states involving two or
three atoms have been produced in experiment [10,11]. In most of the cavity
QED experiments, the cavity field is required to be initially in a pure
state. Previous research show that the microscopic nature of the field is
essential for entangling two or more atoms.

In this paper, we show that maximally entangled states for multiple atoms
can be produced via resonant interaction with a cavity field with a
macroscopic photon number, showing that a macroscopic system can also act as
the entanglement catalyst. Secondly, we show that the entanglement is
insensitive to thermal photons. Under certain conditions the atoms are
disentangled with the cavity field, which is distinguished from the previous
work showing that atom-field entanglement always arises when the field is
initially in a thermal state [12]. Thirdly, the atoms are resonant with the
cavity mode and thus the entanglement speed is very high. More strikingly,
the entanglement speed is independent of both the number of atoms and the
mean photon-number of the thermal field, and high entanglement can appear
even there exist many thermal photons, which is in contrast with the
previous work [13]. Finally, we show that a phase gate between two atoms can
be produced with a thermal field, providing a new prospect for quantum
information processing in a nonzero temperature environment.

Suppose that the single-mode cavity field is initially in the thermal state
\begin{equation}
\rho _{th}=\frac 1{\pi \stackrel{-}{n}}\int e^{-\left| \alpha \right| ^2/%
\stackrel{-}{n}}\left| \alpha \right\rangle \left\langle \alpha \right|
d^2\alpha ,
\end{equation}
where $\stackrel{-}{n}_{th}=1/(e^{\hbar \omega /k_BT}-1)$ is the mean
photon-number of the thermal field. We first displace the cavity field by an
amount $\alpha $, leading to the density operator $D(\alpha )\rho
_{th}D^{+}(\alpha )$, with $D(\alpha )$ being the displacement operator. We
here assume that $\alpha $ is a complex number, i.e., $\alpha =re^{-i\varphi
}$. The displacement can be achieved by injecting the cavity a coherent
field generated by a source [14]. We consider the resonant interaction of N
identical two-level atoms with a single-mode cavity field. In the
rotating-wave approximation, the Hamiltonian is (assuming $\hbar =1$)
\begin{equation}
H=\sum_{j=1}^Ng(a^{+}S_j^{-}+aS_j^{+}),
\end{equation}
where $S_j^{+}=\left| e_j\right\rangle \left\langle g_j\right| $, $%
S_j^{-}=\left| g_j\right\rangle \left\langle e_j\right| $, with $\left|
e_j\right\rangle $ and $\left| g_j\right\rangle $ being the excited and
ground states of the jth atom, $a^{+}$ and $a$ are the creation and
annihilation operators for the cavity mode, and $g$ is the atom-cavity
coupling strength. Suppose that the atoms are initially in the state $\left|
\phi _0\right\rangle $. Then the initial density operator for the whole
system is $\rho _0=D(\alpha )\left| \phi _0\right\rangle \left\langle \phi
_0\right| \otimes \rho _{th}D^{+}(\alpha )$. The evolution operator of the
system is given by $U(t)=e^{-iHt}$. After an interaction time the system
evolves to $\rho =U(t)\rho _0U^{+}(t)$.

We can rewrite the evolution of the density operator as
\begin{equation}
\rho =D(\alpha )U_d(t)\left| \phi _0\right\rangle \left\langle \phi
_0\right| \otimes \rho _{th}U_d^{+}(t)D^{+}(\alpha ),
\end{equation}
where
\begin{eqnarray}
U_d(t) &=&D^{+}(\alpha )U(t)D(\alpha ) \\
&=&e^{-iH_dt},  \nonumber
\end{eqnarray}
where
\begin{equation}
H_d=\sum_{j=1}^Ng[(a^{+}+\alpha ^{*})S_j^{-}+(a+\alpha )S_j^{+}].
\end{equation}
Define the new atomic basis [15,16]

\begin{equation}
\left| +_{j,\varphi }\right\rangle =\frac 1{\sqrt{2}}(\left|
e_j\right\rangle +e^{i\varphi }\left| g_j\right\rangle ),\text{ }\left|
-_{j,\varphi }\right\rangle =\frac 1{\sqrt{2}}(\left| e_j\right\rangle
-e^{i\varphi }\left| g_j\right\rangle ).
\end{equation}
Then we can rewrite $H_d$ as
\begin{eqnarray}
H_d&=&\sum_{j=1}^N\{\frac g2[e^{-i\varphi }a^{+}(2\sigma
_{z,j,\varphi }+\sigma _{j,\varphi }^{+}-\sigma _{j,\varphi
}^{-})\cr&&+e^{i\varphi }a(2\sigma _{z,j,\varphi }+\sigma
_{j,\varphi }^{-}-\sigma _{j,\varphi }^{+})]+2\Omega \sigma
_{z,j,\varphi }\},
\end{eqnarray}
where $\sigma _{z,j,\varphi }=\frac 12(\left| +_{j,\varphi
}\right\rangle \left\langle +_{j,\varphi }\right| -\left|
-_{j,\varphi }\right\rangle \left\langle -_{j,\varphi }\right| ),$
$\sigma _{j,\varphi }^{+}=\left| +_{j,\varphi }\right\rangle
\left\langle -_{j,\varphi }\right| $ , $\sigma _{j,\varphi
}^{-}=\left| -_{j,\varphi }\right\rangle \left\langle +_{j,\varphi
}\right| ,$ and $\Omega =rg$. We can rewrite $U_d(t)$ as
\begin{equation}
U_d(t)=e^{-i2\Omega \sigma _{z,j,\varphi }t}e^{-iH_it},
\end{equation}
where
\begin{eqnarray}
H_i&=&\sum_{j=1}^N\frac g2[e^{-i\varphi }a^{+}(2\sigma _{z,j,\varphi
}+e^{i2\Omega t}\sigma _{j,\varphi }^{+}-e^{-i2\Omega t}\sigma
_{j,\varphi }^{-})\cr&&+e^{i\varphi }a(2\sigma _{z,j,\varphi
}+e^{-i2\Omega t}\sigma _{j,\varphi }^{-}-e^{i2\Omega t}\sigma
_{j,\varphi }^{+})]
\end{eqnarray}
Assuming that $\Omega \gg g,$ we can neglect the terms oscillating
fast. Then $H_i$ reduces to
\begin{equation}
H_i=g(e^{-i\varphi }a^{+}+e^{i\varphi }a)\sigma _{z,\varphi },
\end{equation}
where
\begin{equation}
\sigma _{z,\varphi }=\sum_{j=1}^N\sigma _{z,j,\varphi }\text{.}
\end{equation}

The Hamiltonian $H_i$ describes a spin-dependent force on the cavity field.
It has been shown that this Hamiltonian can be obtained in the ion trap
[17,18]. In this case, a collective vibrational mode acts as the bosonic
system and the internal degrees of freedom of the ions correspond to the
spin system. The spin-dependent force has been used to generate
Schr\"odinger cat states [19] and implement two-qubit phase gates [20] in
ion trap experiments. Milburn et al. [17] have proposed a scheme for
realizing multi-qubit gates via sequent applications of the Hamiltonian with
variable parameter $\varphi $. In the ion trap, $\varphi $ is adjustable via
the phases of driving lasers resonant with the sideband transitions. The aim
of the following section is to show that sequent spin-dependent forces with
controllable parameter $\varphi $ can be achieved in the atom-cavity system
by applying a sequence of displacements and atomic rotations interspersed
between periods of evolution of the system. The corresponding spin-dependent
displacement along a close path in phase space produces a spin-dependent
phase, which can be used to generate Greenberger-Horne-Zeilinger (GHZ)
states and implement two-qubit phase gates.

Define the symmetrical state $\left| \Phi _{k,\varphi }\right\rangle $ with
k atoms being in the state $\left| -_{j,\varphi }\right\rangle $, i.e., the
well known Dicke state [21]. Applying the collective atomic operator $\sigma
_{z,\varphi }$ to the Dicke state $\left| \Phi _{k,\varphi }\right\rangle $,
we obtain

\begin{equation}
\sigma _{z,\varphi }\left| \Phi _{k,\varphi }\right\rangle =(N/2-k)\left|
\Phi _{k,\varphi }\right\rangle .
\end{equation}
We first assume that $\varphi =0$, i.e., $\alpha =r$. We now assume that
each atom is initially in the state $\left| e_j\right\rangle .$ $\left|
e_j\right\rangle $ can be rewritten as

\begin{equation}
\left| e_j\right\rangle =\frac 1{\sqrt{2}}(\left| +_{j,0}\right\rangle
+\left| -_{j,0}\right\rangle ).
\end{equation}
Then the initial state for the N atoms can be written as a Bloch state [22]
\begin{equation}
\left| \phi _0\right\rangle =\frac 1{\sqrt{2^N}}\sum_{k=0}^N\left(
\begin{array}{c}
k \\
N
\end{array}
\right) ^{1/2}\left| \Phi _{k,0}\right\rangle .
\end{equation}
Using Eqs.(3), (8), (10), (12), and (14), we obtain evolution of the system
after an interaction time $\tau $

\begin{eqnarray}
\rho _1 &=&\frac 1{2^N}\sum_{k=0}^N\sum_{k^{^{\prime
}}=0}^Ne^{2i(k-k^{^{\prime }})\Omega \tau }\left(
\begin{array}{c}
k \\
N
\end{array}
\right) ^{1/2}\left(
\begin{array}{c}
k^{^{\prime }} \\
N
\end{array}
\right) ^{1/2}\left| \Phi _{k,0}\right\rangle \left\langle \Phi
_{k^{^{\prime }},0}\right| \\
&&\otimes D(r)D[-i(N/2-k)g\tau ]\rho _{th}D^{+}[-i(N/2-k^{^{\prime }})g\tau
]D^{+}(r),  \nonumber
\end{eqnarray}
The resonant interaction of the atom with the strongly displaced thermal
field results in the spin-dependent displacement operator on the cavity mode.

We then displace the cavity mode by an amount $-r+ir$ and perform the
rotation $\left| g_j\right\rangle \rightarrow i\left| g_j\right\rangle $,
which leads to
\begin{eqnarray}
\rho _1^{^{\prime }} &=&\frac 1{2^N}\sum_{k=0}^N\sum_{k^{^{\prime
}}=0}^Ne^{2i(k-k^{^{\prime }})\Omega \tau }\left(
\begin{array}{c}
k \\
N
\end{array}
\right) ^{1/2}\left| \Phi _{k,\pi /2}\right\rangle \left\langle \Phi
_{k^{^{\prime }},\pi /2}\right| \\
&&\ \otimes D(ir)D[-i(N/2-k)g\tau ]\rho _{th}D^{+}[-i(N/2-k^{^{\prime
}})g\tau ]D^{+}(ir).  \nonumber
\end{eqnarray}
After an interaction time $\tau ,$ we obtain
\begin{eqnarray}
\rho _2 &=&\frac 1{2^N}\sum_{k=0}^N\sum_{k^{^{\prime
}}=0}^Ne^{4i(k-k^{^{\prime }})\Omega \tau }\left(
\begin{array}{c}
k \\
N
\end{array}
\right) ^{1/2}\left| \Phi _{k,\pi /2}\right\rangle \left\langle \Phi
_{k^{^{\prime }},\pi /2}\right| \\
&&\ \ \otimes D(ir)D[2(N/2-k)g\tau ]D[-i(N/2-k)g\tau ]\rho _{th}  \nonumber
\\
&&D^{+}[-i(N/2-k^{^{\prime }})g\tau ]D^{+}[2(N/2-k)g\tau ]D^{+}(ir).
\nonumber
\end{eqnarray}
After the field displacement and atomic rotation, the resonant interaction
yields a second spin-dependent displacement perpendicular to the first one
in phase space. This is due to the fact that the total displacement before
the second resonant interaction is just perpendicular to that before the
first resonant interaction.

We repeat the procedure for two more times. During the two cycles the
displacements are $-r-ir$ and $r-ir$, respectively. The final state of the
system is
\begin{eqnarray}
\rho _f &=&\frac 1{2^N}\sum_{k=0}^N\sum_{k^{^{\prime
}}=0}^Ne^{8i(k-k^{^{\prime }})\Omega \tau }\left(
\begin{array}{c}
k \\
N
\end{array}
\right) ^{1/2}\left| \Phi _{k,-\pi /2}\right\rangle \left\langle \Phi
_{k,-\pi /2}\right| \\
&&\ \ \ \otimes D(-ir)D[-2(N/2-k)g\tau ]D[i(N/2-k)g\tau ]D[2(N/2-k)g\tau
]D[-i(N/2-k)g\tau ]\rho _{th}  \nonumber \\
&&D^{+}[-i(N/2-k^{^{\prime }})g\tau ]D^{+}[2(N/2-k^{^{\prime }})g\tau
]D^{+}[i(N/2-k^{^{\prime }})g\tau ]D^{+}[-2(N/2-k^{^{\prime }})g\tau
]D^{+}(-ir).  \nonumber
\end{eqnarray}
We can rewrite $\rho _f$ as
\begin{equation}
\rho _f=\left| \phi _f\right\rangle \left\langle \phi _f\right| \otimes
D(-ir)\rho _{th}D^{+}(-ir)
\end{equation}
where
\begin{eqnarray}
\left| \phi _f\right\rangle &=&\frac 1{\sqrt{2^N}}\sum_{k=0}^Ne^{8ik\Omega
\tau }e^{2i[(N/2-k)g\tau ]^2}\left(
\begin{array}{c}
k \\
N
\end{array}
\right) ^{1/2}\left| \Phi _{k,-\pi /2}\right\rangle \\
&=&\frac 1{\sqrt{2^N}}\sum_{k=0}^Ne^{8ik\Omega \tau -2ikN(g\tau
)^2}e^{2ik^2(g\tau )^2}\left(
\begin{array}{c}
k \\
N
\end{array}
\right) ^{1/2}\left| \Phi _{k,-\pi /2}\right\rangle .  \nonumber
\end{eqnarray}
We here have discarded the common phase factor $e^{iN^2(g\tau )^2/2}$. With
the choice $2(g\tau )^2=\pi /2$ we obtain [23,24]
\begin{eqnarray}
\left| \phi _f\right\rangle &=&\frac 1{\sqrt{2^{N+1}}}[e^{i\pi
/4}\prod_{j=1}^N(\left| +_{j,-\pi /2}\right\rangle +e^{8i\Omega \tau -iN\pi
/2}\left| -_{j,-\pi /2}\right\rangle ) \\
&&+e^{-i\pi /4}\prod_{j=1}^N(\left| +_{j,-\pi /2}\right\rangle -e^{8i\Omega
\tau -iN\pi /2}\left| -_{j,-\pi /2}\right\rangle )].  \nonumber
\end{eqnarray}
Since the state $(\left| +_{j,-\pi /2}\right\rangle +e^{8i\Omega \tau -iN\pi
/2}\left| -_{j,-\pi /2}\right\rangle )/\sqrt{2}$ is orthogonal to $(\left|
+_{j,-\pi /2}\right\rangle -e^{8i\Omega \tau -iN\pi /2}\left| -_{j,-\pi
/2}\right\rangle )/\sqrt{2}$, $\left| \phi _f\right\rangle $ is a N-particle
maximally entangled state, or a GHZ state [2]. The average photon-number of
the displaced thermal state depends upon the amount of the initial
displacement: $\stackrel{-}{n}=$ $\stackrel{-}{n}_{th}+\left| \alpha \right|
^2$. The entanglement persists in the classical limit $\left| \alpha \right|
^2\rightarrow \infty $.

The entanglement speed is independent of both the number of atoms and the
mean photon-number of the thermal field. The strongly displaced JCM
evolution operator produces a displacement conditional on the atomic state.
The cavity field is displaced along the sides of a square, whose length
depends upon the state of the atomic system. After the operation, the atomic
system is disentangled with the cavity field, but acquires a phase
conditional on the displace path [17,18,20], leading to the entanglement.
The macroscopic thermal field acts as the entanglement catalyst.

We note the idea can be generalized to realize geometric phase gates for two
atoms with a thermal field. For the two-atom case, the above mentioned
displacements, rotations, and resonant interactions leads to the
transformation:
\begin{eqnarray}
\left| +_{1,0}\right\rangle \left| +_{2,0}\right\rangle &\rightarrow
&e^{i2(g\tau )^2}\left| +_{1,-\pi /2}\right\rangle \left| +_{2,-\pi
/2}\right\rangle ,\text{ } \\
\left| +_{1,0}\right\rangle \left| -_{2,0}\right\rangle &\rightarrow &\left|
+_{1,-\pi /2}\right\rangle \left| -_{2,-\pi /2}\right\rangle ,  \nonumber \\
\left| -_{1,0}\right\rangle \left| +_{2,0}\right\rangle &\rightarrow &\left|
-_{1,-\pi /2}\right\rangle \left| +_{2,-\pi /2}\right\rangle ,\text{ }
\nonumber \\
\left| -_{1,0}\right\rangle \left| -_{2,0}\right\rangle &\rightarrow
&e^{i2(g\tau )^2}\left| -_{1,-\pi /2}\right\rangle \left| -_{2,-\pi
/2}\right\rangle .  \nonumber
\end{eqnarray}
We here have discarded the trival single-qubit phase shifts, which can be
absorbed into next single-qubit operations. Setting $2(g\tau )^2=\pi /2$ and
performing the rotation $\left| g_j\right\rangle \rightarrow i\left|
g_j\right\rangle $ we obtain the phase gate
\begin{eqnarray}
\left| +_{1,0}\right\rangle \left| +_{2,0}\right\rangle &\rightarrow
&i\left| +_{1,0}\right\rangle \left| +_{2,0}\right\rangle ,\text{ } \\
\left| +_{1,0}\right\rangle \left| -_{2,0}\right\rangle &\rightarrow &\left|
+_{1,0}\right\rangle \left| -_{2,0}\right\rangle ,  \nonumber \\
\left| -_{1,0}\right\rangle \left| +_{2,0}\right\rangle &\rightarrow &\left|
-_{1,0}\right\rangle \left| +_{2,0}\right\rangle ,\text{ }  \nonumber \\
\left| -_{1,0}\right\rangle \left| -_{2,0}\right\rangle &\rightarrow
&i\left| -_{1,0}\right\rangle \left| -_{2,0}\right\rangle .  \nonumber
\end{eqnarray}
The combination of this gate and the single-qubit phase shifts $\left|
+_{1,0}\right\rangle \rightarrow -i\left| +_{1,0}\right\rangle $ and $\left|
-_{2,0}\right\rangle \rightarrow i\left| -_{2,0}\right\rangle $ corresponds
to two-qubit $\pi $-phase gate.

We now show how the gate is robust against parameter fluctuations. Suppose
that the two atoms are initially in the state

\begin{equation}
\left| \phi _0\right\rangle =\frac 12(\left| +_{1,0}\right\rangle +\left|
-_{1,0}\right\rangle )(\left| +_{2,0}\right\rangle +\left|
-_{2,0}\right\rangle ).
\end{equation}
The phase gate of Eq. (23) produces the maximally entangled state
\begin{equation}
\left| \phi _f\right\rangle =\frac 12[\left| +_{1,0}\right\rangle (i\left|
+_{2,0}\right\rangle +\left| -_{2,0}\right\rangle )+\left|
-_{1,0}\right\rangle (\left| +_{2,0}\right\rangle +i\left|
-_{2,0}\right\rangle )].
\end{equation}
If the condition $2(g\tau )^2=\pi /2$ is not exactly satisfied, the final
state is
\begin{equation}
\left| \phi _f^{^{\prime }}\right\rangle =\frac 12[\left|
+_{1,0}\right\rangle (e^{i2(g\tau )^2}\left| +_{2,0}\right\rangle +\left|
-_{2,0}\right\rangle )+\left| -_{1,0}\right\rangle (\left|
+_{2,0}\right\rangle +e^{i2(g\tau )^2}\left| -_{2,0}\right\rangle )].
\end{equation}
The fidelity is given by
\begin{equation}
F=\left| \left\langle \phi _f\right| \left. \phi _f^{^{\prime
}}\right\rangle \right| ^2=\frac 14\{1+\sin [2(g\tau )^2]\}^2+\frac 14\cos
^2[2(g\tau )^2].
\end{equation}
Set $2(g\tau )^2=0.55\pi $. Then the fidelity is about $0.99$.

In microwave cavity QED experiments [10,25], two or more atoms are
simultaneously sent through a cavity. For the Rydberg atoms with principal
quantum numbers 50 and 51, the radiative time is $T_r=3\times 10^{-2}s$, and
the coupling constant is $g=2\pi \times 25kHz$ [10]. Thus the interaction
times of atoms with the cavity field are $t=4\tau =2\sqrt{\pi }/g=2.26\times
10^{-5}s$. In the case $N=3$ the decoherence time of the atomic system is $%
T_r^{^{\prime }}=T_r/3=1\times 10^{-2}s$. The decoherence time of the
superposition of different components in the nonzero temperature heat bath
is $T_c^{^{\prime }}=T_c/(1+2\stackrel{-}{n}_{th})d^2$, where $d$ is the
distance between the components in phase space. During the interaction, the
distance between the coherent components is on the order of $g\tau \sim 1$.
Very recently, a ultrahigh fineness Fabry-Perot resonator with a damping
time $T_c=0.13s$ has been built [25]. Set $\stackrel{-}{n}_{th}=5$. Then,
the decoherence time for the cavity field is about $T_c^{^{\prime }}\sim
6.19\times 10^{-3}s$. The infidelity induced by the decoherence is about $%
t/T_r+t/T_c^{^{\prime }}=0.591\times 10^{-2}$.

In conclusion, we have shown that maximally entangled states for multiple
atoms can be induced by a cavity field initially in a thermal state. The
entanglement appears in the macroscopic limit. The time for the appearance
of maximal entanglement is independent of both the number of atoms and the
mean photon-number of the thermal field. The quantum phase gate for two
atoms can also be produced via interaction with the displaced thermal field.
The macroscopic thermal field acts as the catalyst for producing
entanglement and quantum information processors.

This work was supported by the National Natural Science Foundation of China
under Grant No. 10674025 and funds from Key Laboratory of Quantum
Information, University of Science and Technology of China.

\end{document}